\documentclass{aastex}

\usepackage{emulateapj5}

\usepackage{natbib} 

\newcommand\ha{{H$\alpha$}}

\newcommand\kms{\:\rm{\,km\,s^{-1}}}
\newcommand\asy{\:\rm{\,arcsec\,yr^{-1}}}
\newcommand\masy{\:\rm{\,mas\:yr^{-1}}}

\newcommand\FLUXARCSEC{\:{\rm ergs\:cm^{-2}\:s^{-1}\:arcsec^{-2}}}

\newcommand\eg{{\it e.g.}}
\newcommand\ie{{\it i.e.}}
\newcommand\etal{{\it et\thinspace al.}\ }

\journalinfo{Submitted to {\it The Astrophysical Journal}.}
\slugcomment{Submitted:  30 July 2002}

\shorttitle{SN~1006 Optical Proper Motions, etc.}
\shortauthors{Winkler, Gupta, \& Long}

\begin{document}


\title{The SN~1006 Remnant: Optical Proper Motions, Deep Imaging,  
Distance, and Brightness at Maximum}


\author{P. Frank Winkler\altaffilmark{1},  Gaurav Gupta\altaffilmark{2}}
\affil{Department of Physics, Middlebury College, Middlebury, VT 05753}
\email{winkler@middlebury.edu}

\and

\author{Knox S. Long\altaffilmark{1}}
\affil{Space Telescope Science Institute, Baltimore MD 21218}

\altaffiltext{1}{Visiting Astronomer, Cerro Tololo Inter-American Observatory.
CTIO is operated by AURA, Inc.\ under contract to the National Science
Foundation.}
\altaffiltext{2}{Present Address: Department of Mechanical and Aerospace Engineering,
Cornell University, Ithaca, NY  14853}


\begin{abstract}
We report the first measurement of proper motions in 
the SN~1006 remnant (G327.6+14.6) based entirely on digital images.  
CCD images from three epochs spanning a period of 11 years are used:
1987 from Las Campanas, and 1991 and 1998 from CTIO.  
The filaments in SN~1006 are non-radiative, appearing only in
the Balmer lines of hydrogen.  Delicate filaments---probably thin 
sheets seen in projection---delineate the shock front along the 
northwest rim of the remnant.  We use the center of curvature of the 
filaments to define a convenient geometry, and integrate along 
short segments to obtain one-dimensional profiles.  These remain 
unchanged, within seeing and statistical errors, from one epoch 
to another.  Measuring the shift in a direction perpendicular to
the filaments, we obtain proper motions of 
$280 \pm 8 \masy$\ along the entire length where the filaments 
are well defined, with little systematic variation along the 
filaments.  In addition to the measurements of the well-defined filaments in 
the NW of SN~1006, we also report very deep \ha\ imaging observations of 
the entire remnant that clearly show very faint emission surrounding 
almost the entire  shell, as well as some diffuse emission regions  
in the (projected) interior.
  
Combining the proper motion measurement with
a recent measurement of the shock velocity based on spectra of the
same filaments by Ghavamian \etal\
leads to a distance of $2.17\pm 0.08\; {\rm kpc}$ to SN~1006.  
Several lines of argument suggest that SN~1006 was a Type Ia event, so 
the improved distance measurement can be combined with  
the peak luminosity for SNe~Ia, as recently determined for events 
in galaxies with 
Cepheid-based distances, to calculate the apparent brightness of the 
spectacular event 
that drew wide attention in the eleventh century.  The result, 
$V_{max} = -7.5 \pm 0.4$, lies squarely in the middle of the wide range of 
estimates based on the historical observations.  

\end{abstract}




\section{Introduction}

Early in the eleventh century, the supernova of 1006 C.E. attracted
wide attention, and it is probably the brightest stellar event in recorded 
human history \citep{ste77}.  Almost a millennium later, the supernova 
remnant (SNR) that this event produced
has gradually emerged as an object of great astrophysical interest.   
The remnant of SN~1006, G327.6+14.6, was first identified through its radio emission by 
\citet{gar65}.  X-ray emission was detected by 
\citet{win76}, and very soon afterward \citet{vdb76} reported faint optical 
filaments associated with SN~1006.   More recent radio maps 
\citep{rey86, mof93} show a somewhat 
barrel-shaped shell, $30\arcmin$\ in diameter and brightest along the NE and SW limbs. 
X-ray images \citep[][hereafter WL97]{wil96, win97a} show a similar structure to 
that seen in the radio, while optical emission is limited almost entirely 
to delicate filaments extending some $15\arcmin$\ along the NW rim of the 
remnant shell---a region that is generally faint in both X-rays and radio.  
The optical spectrum shows only Balmer lines of hydrogen \citep{sch78}, 
indicative of a nonradiative shock  propagating into partially neutral 
interstellar material \citep[\eg][]{mck80}.  

\begin{table*}[t]


\center{
\scriptsize
{\scshape  
\vspace {0.02in}
Table 1 \\

Imaging Observations of SN~1006.} \\


\begin{tabular}{lcccccccl}
\\
\tableline \tableline
\\
\vspace {0.02in}
 &  &  & Scale & \multicolumn{2}{c}{Filter} &   
Exposure  & Seeing &  \\
 
\vspace {0.05in}
Date & Telescope   & CCD   &  
($\arcsec\;{\rm pixel}^{-1}$) &
$\lambda_0$\ (\AA) &$\Delta\lambda$\ (\AA)  & 
(s) & (FWHM) &  Observers 
\\


\tableline  \\

1987 Apr 26 & LCO 2.5-m & TI $800\times800$ & 0.41 & 6572  & 52 & $3 \times 1000 $ & 
1\farcs0 & Long \& Blair$^a$   \\
1991 Apr 19 &CTIO 4-m &Tek $1024\times 1024$ & 0.47 &6563 & 30 & 600  &
1\farcs1 & Winkler \& Long\\
\vspace {0.05in}
1998 Jun 23 &CTIO 0.9-m &Tek $2048\times 2048$ & 0.40 &6566  & 24 & $9 \times 1000 $ &
1\farcs1 & Winkler \& Long\\

\tableline
\tiny

\\
\multicolumn{9}{l}{$^a$William P. Blair, Johns Hopkins University.}

\end{tabular}
}

\end{table*}

Renewed interest in the  SN~1006 remnant was stimulated with 
the discovery by \citet{koy95}, based on {\it ASCA} data, 
that the strongest X-ray emission, from the NE and 
SW limbs, has a hard, featureless, non-thermal spectrum 
characteristic of synchrotron radiation, while the X-rays from the 
interior and from the NW and SE limbs are much softer and are thermal in 
character.  These data strongly suggested that in the NE and SW, electrons are accelerated to 
TeV energies by the first-order Fermi mechanism in the supernova shock.  
Furthermore, since the same mechanism will accelerate nuclei as well, the 
\citet{koy95} data provided the first solid observational evidence that supernova shocks 
produce cosmic rays.  The supernova-cosmic ray connection was solidified 
with the demonstration of detailed correspondence between X-ray and radio 
morphology along the NE and SW limbs of SN~1006 (WL97) and 
especially by the discovery of $\gamma$-ray emission from the NE limb of 
SN~1006 by \citet{tan98}.

The optical filaments of SN~1006 provide a rich opportunity to investigate 
the nature of the nonradiative shock and the environment into which it 
propagates.   
The Balmer line profile consists of two components: a broad component that 
results from neutral atoms that penetrate the shock and then undergo 
charge exchange with hot, post-shock protons to produce a population of hot 
neutrals, and a narrow component from 
simple collisional excitation of the ambient neutral atoms \citep{che78, 
che80}.  \citet{gha02} present detailed spectra and models  that lead to a precise 
measurement of the shock velocity in the NW region of SN~1006: $v_s = 
2890\pm 100\ \kms$.  


The present paper presents a measurement of the proper motion of the NW 
filaments of SN~1006 with a precision of $\sim 2\%$, comparable to that of 
the recent shock velocity measurement.  
The geometry of the NW filaments---apparently thin sheets seen 
almost exactly edge-on---is ideally suited to measuring the remnant's 
expansion.  
Proper motion of the optical filaments in SN~1006 was first reported only 
five years after their discovery \citep{hes81}, but a reasonably 
precise measurement had to wait for a longer baseline. \citet[hereafter LBV88]{lon88} 
combined first-epoch photographic plates and CCD images  taken 11 years 
later to obtain a measurement of $0.30 \pm 0.04\asy$.  We use the same 
1987-epoch CCD image used by LBV88, together with subsequent CCD images from 
epochs 1991 and 1998, to obtain a much more precise measurement---the first 
relying entirely on digital images.   Section 2 of the paper describes 
the observations themselves.  Section 3 discusses two different algorithms  
we have employed to measure the filamentary motion and presents the 
results.  Section 4 presents new, deep observations of very faint 
\ha\ emission that outlines the entire shell of the SN~1006 remnant, as 
first reported by WL97.  In the final three sections we discuss some
implications of the present measurements, including a comparison with 
proper motion measurements carried out in other regimes of the 
electromagnetic spectrum.  The combination of our new proper-motion 
measurements and the shock-velocity determined by \citet{gha02} leads 
directly to a distance of $2.17\pm 0.08\; {\rm kpc}$\  pc to SN~1006.  
This distance, together with recent determinations of the peak absolute 
magnitudes for Type Ia supernovae, provides an estimate of the  
apparent brightness of SN~1006 at maximum that is significantly more precise than 
earlier ones.

\section{Observations}

Narrow-band imaging observations of SN~1006 in \ha\ were carried out
at three epochs, 1987, 1991, and 1998, using CCDs on telescopes
at Las Campanas and CTIO.  The observational details are summarized
in Table 1.  All three sets of observations were carried out under
good seeing conditions; individual stars show profiles with FWHM near 
1\arcsec\@. 
The first-epoch image for the
current study is identical with the one used for the second epoch 
by LBV88 and appearing in Figures 1 and 3a of that paper, and  
the 1991-epoch image from the Blanco 4-m telescope 
appears as part of the mosaic shown as Figure 2 in WL97.    
The data were reduced using standard 
IRAF\footnote{IRAF is distributed by the National Optical Astronomy 
Observatories, which is operated by the  AURA, Inc. under cooperative 
agreement with the National Science Foundation.} 
procedures of bias subtraction and flat fielding.  

\section{Proper Motion Measurement}
The initial step in measuring filamentary proper motions is to align images from 
different epochs so they are as perfectly registered as possible, using 
the stars as fiducial markers.  Then the filamentary motions are 
apparent as position shifts from one epoch  to another.   
In their measurement, LBV88 simply used cursor readings at several points 
along the filament to compare the positions at two epochs.  Since their
first-epoch image came from a digitized photographic plate, they found  
that this technique was as accurate as any other; nevertheless, this 
limited the precision of their measurement to $\sim 1\; {\rm pixel} = 
0\farcs 4$\ or $\pm 0.037\; {\rm arcsec\; yr^{-1}}$\ over their 11-year 
baseline.  

With multiple epochs of CCD data, we can take a more sophisticated 
approach to obtain measurements with a precision of a small 
fraction of a pixel.  The basic idea is first to obtain a one-dimensional profile 
in a direction perpendicular to a short segment of filament at multiple epochs.  Then we 
shift the profile from one epoch relative to that at another and 
measure how well they match as a function of the shift.  The shift that  
gives the best match indicates the proper motion between the two 
epochs for that segment.  This section describes the steps in some detail.

\subsection{Image Registration}
We first aligned the images from 
all three epochs on a common coordinate system using tasks in the IRAF 
``imcoords'' and ``immatch'' packages.  In order to achieve optimum resolution, we transformed 
all the images to a standard coordinate system at a scale of exactly 
$0\farcs 10\; {\rm pixel}^{-1}$, subsampling the original images by about a 
factor of 4.  We chose the 1998  
epoch as the reference image, since it has a 13\farcm7 field of view,  widest 
of the three epochs in the set, and minimal distortion across the field. 
We obtained plate solutions for  each of the nine individual 1998 frames, all of which were taken 
on somewhat different centers, using stars from the UCAC1 catalog 
\citep{ucac1_00}.  The simplest 4-parameter solution ($x$- and 
$y$-translation, rotation, and pixel scale) gave excellent fits.  We then 
transformed the individual frames to our adopted standard coordinate 
system, using simple linear interpolation for the subsampling, and 
combined them into a single reference image.  
 
We then transformed the images from the 1987 and 1991 epochs to match the 
reference image.  Here the transformation was less straightforward, due to 
the lack of flatness that characterized the TI chip used in 1987 and the 
curvature of the Blanco 4m prime-focus camera used in 1991.    
For fiducial points, we used a  set of over 100
stars, located in a band straddling the NW filaments, that are unsaturated in all three 
images.  After allowing quadratic distortion terms, we obtained   
good fits, with RMS residuals of $ 0\farcs 06$\ and  $0\farcs 04$\ 
for the 1987 and 1991 epoch images, respectively.  (An exception is
the extreme western edge of the 1987 image, which shows larger 
distortions where the TI chip curvature was the worst and the stars were 
poorly focused.  This region was not used in our analysis.)  

We note that the rebinned images have somewhat lower (by a factor of about 1.4) sky 
noise than the original ones, for which the sky noise was completely 
consistent with Poisson statistics.  
This results because of correlations introduced in the image transformation 
process  \citep{fru02} and averaging of flux across multiple pixels.
In our subsequent 
analysis, in which we use uncertainties in the pixel values as the basis 
for calculating $\chi^2$\ values, we have adjusted these uncertainties from 
their purely Poisson-statistical values to reflect the averaging effect 
and to be consistent with the actual distribution in the rebinned, 
registered images. 

Once the images from the three epochs have been registered, the proper motion 
of the NW filaments is apparent.  This is illustrated dramatically in 
Figure 1: the left panel shows a section of the 1998 image, while the 
right panel shows the difference between the 1998 and 1987 images.

\begin{figure*}
\epsscale{1.0}
\plotone{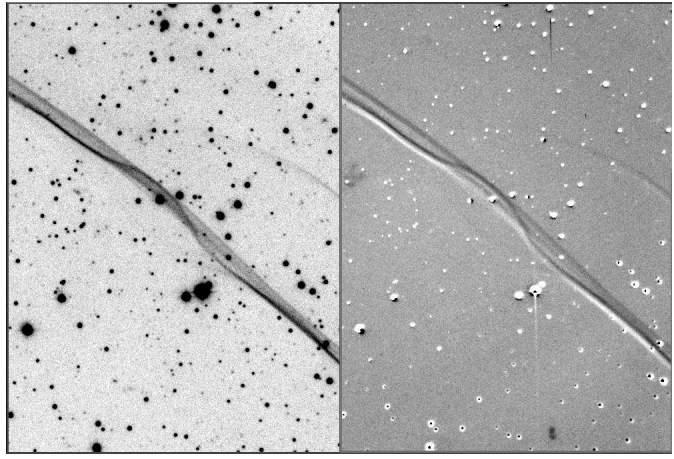}
\caption{
(Left) A portion of the NW rim of the SN~1006 remnant in \ha\ showing the
prominent filaments at epoch 1998.   (Right)   
The same image with one from epoch 1987 
subtracted,  so that at the earlier epoch the filaments appear white, and 
at the later one, black.
}
\end{figure*}

\begin{figure*}
\epsscale{1.0}
\plotone{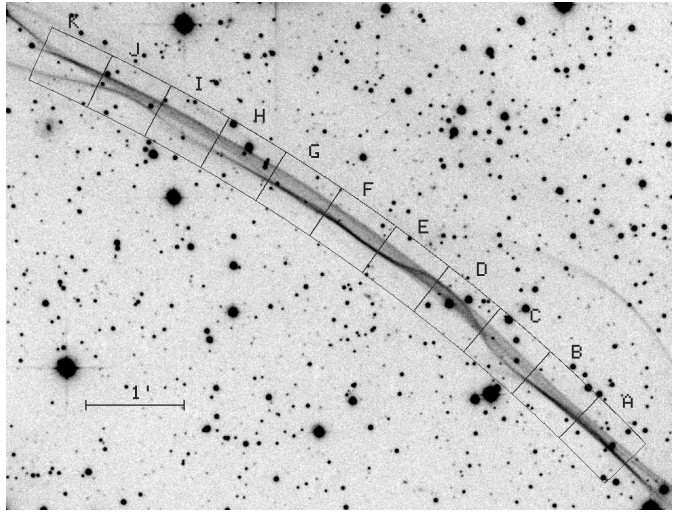}
\caption{
The 1998-epoch \ha\ image of the NW filaments in SN~1006, showing the 
11 sectors (A--K) for which radial profiles and proper motions were 
measured.  (For the 1987-epoch image, only sectors B--I were usable.)
}
\end{figure*}

\subsection{One-dimensional Profiles}

We sought to measure one-dimensional profiles across the narrow width of 
the filaments, \ie, along their direction of motion.  
In order to get profiles with both high angular resolution across the 
filaments and reasonable signal-to-noise, we must integrate for some 
distance along them.  Fortunately, the geometry of the NW filaments 
is very favorable; their crisp, well-defined portion lies 
very nearly along the arc of a circle, centered at 
$\alpha(2000) = 15^{\rm h}03^{\rm m}16\farcs2,\ \delta (2000) = 
-42\degr 00\arcmin 30\arcsec$\ (WL97).  (This is {\it not} the center of the 
SNR shell, but rather the center of curvature for the brightest optical filaments.)  From 
this center we defined 11 sectors, each $2\degr$\ in azimuth, and took a  
radial profile for the portion of the filament within each sector.  Each 
bin of a profile represents a 
very thin slice of width 0\farcs10 integrated along 
$2\degr$\ in azimuth, an arc length of about 40 arcsec.  
The geometry is illustrated in Fig. 2.  

Before measuring the radial profiles for the filament, we first had to 
remove the stars from the region of interest; otherwise, they would 
contaminate the profiles and confuse the measurement.  After the images 
had been aligned, stars within the profile region were subtracted using 
DAOPHOT techniques, followed by manual editing.  Then the radial profiles were 
calculated using the IRAF/PROS task ``imcnts'' to measure the 
average counts per pixel along each of 350  0\farcs 1-wide slices.  
Profiles were obtained for all eleven sectors at 
the 1991 and 1998 epochs.  With the 1987-epoch image, however,
the smaller field of view and distortions near 
the western edge limited us to using profiles for only eight of the sectors. 
A common scaling factor was applied to all the profiles 
from each epoch to account for differences in exposure time and 
sensitivity among the observations from different instruments at the three 
epochs.  For each profile we also calculated the 
uncertainty in each of the bins based on Poisson statistics from the 
original images (using the known gain and read-out noise), adjusted downward 
slightly to reflect the averaging that accompanied rebinning (see previous 
section).  As an important check on 
this procedure, we found that the actual measured rms dispersion for the 
sky portion of the profiles (\ie, in regions inside or outside the 
filaments, where the counts per pixel 
appeared to be constant, on average) agreed well  with the uncertainty values 
we had calculated.  Scaled profiles for one sector for epochs 1987 
and 1998 are shown in Fig. 3.

\subsection{Measuring Shifts between Epochs}
Once scaled profiles like the ones shown in Fig. 3 have been calculated, 
it remains merely to shift the profile from one epoch outward in the radial 
direction to achieve the best match with the profile at a later epoch.
Since the profiles are well-defined and have high signal-to-noise, 
simply sliding profiles from different epochs back  and forth and matching 
them by eye is sufficient to measure the shift to within a few tenths of 
an arcsec; however, digital techniques make more precise measurement possible.  
We have applied two different 
techniques, outlined below, which give virtually identical results.

\subsubsection{Method I:  Correlation Analysis} 
The problem of shifting a one-dimensional image by a variable offset to 
achieve the best match to a template image occurs  
often in spectroscopy; \eg, in calculating the redshifts of distant 
galaxies.  \citet{ton79} have presented a procedure based on 
correlation analysis for calculating the optimum redshift of a spectrum. 
We followed a virtually identical procedure to obtain the pixel shift 
of both the 1987 and 1991 profiles relative to the corresponding profiles 
in the 1998 reference image.   This method is quite straightforward to implement  
using the IRAF task ``fxcor'' to obtain a correlation in terms of profile bins, 
but in our case it does not lead to 
reliable estimates for the uncertainty in the measured value for pixel 
shift between epochs.

\begin{figure*}[t]
\epsscale{0.75}
\plotone{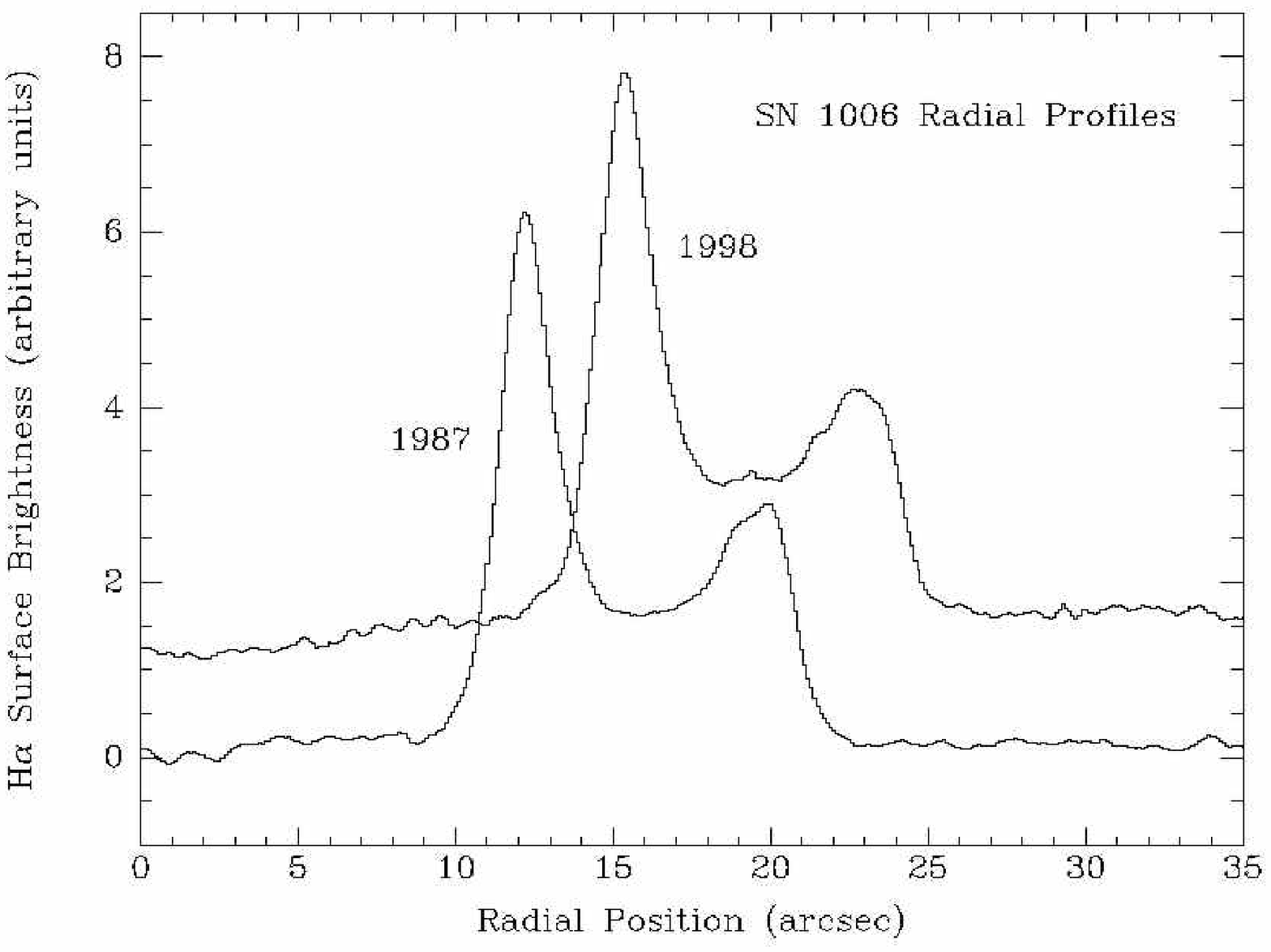}
\caption{
Radial profiles of a portion of the NW filament of SN~1006 
(sector F of Fig. 2) at 
epochs 1987 and 1998.  (The 1998 profile is shifted upward for 
clarity.)  For this and other sectors the profile shapes are virtually identical 
at different epochs except for the obvious radial displacement.
}
\end{figure*}

\subsubsection{Method II:  Minimum $\chi^2$\  Analysis}
We have also used an independent approach 
based on calculating directly the $\chi^2$\ probability 
that two observed profiles from different epochs represent the same 
underlying physical profile, shifted by an amount treated as a parameter 
in the fit.  
Let $x_1(k)$\ and $x_2(k)$\ represent the observed profiles in discrete 
bins ($k$) at epochs 1 and 2, and $\sigma_1(k)$\ and $\sigma_2(k)$\ 
represent the uncertainties at each bin.  
These uncertainties are traceable to  Poisson statistics and the 
read noise of the chips, with some adjustment as discussed in Section 3.2.  
If  $x_1(k)$\ and $x_2(k)$\ are simply two independent samples of the same 
underlying physical distribution, then we expect 
a $\chi^2$-distribution with $n$\ degrees of freedom for
\begin{equation}
\chi ^2=\sum\limits_{k=1}^n {{{\left( {x_1(k)-x_2(k)} \right)^2} \over 
{\sigma _1^2(k)+\sigma _2^2(k)}}},
\end{equation}
where $n$\ is the 
number of bins in the profile \citep[\eg,][]{bev92}.
For the case in which one profile is offset relative to another, we 
introduce an offset parameter, $l$, in the bin index and calculate
$\chi^2$\ as a function of $l$,
\begin{equation}
\chi ^2(l)=\sum\limits_{k=1}^n {{{\left( {x_1(k+l)-x_2(k)} \right)^2} \over 
{\sigma _1^2(k+l)+\sigma _2^2(k)}}}.
\end{equation}
The minimum for $\chi^2(l)$\ occurs for the value of $l$\ at which the two 
profiles best match.  Furthermore, since this represents a fit 
with  $l$\ as a single ``interesting'' parameter, the $1\, \sigma$\ 
uncertainty limits will occur for values of  $l$\ that give 
$\chi^2_{min} + 1$\ \citep{lam76,avn76}.  And finally, the reduced $\chi^2$\ value,
$\chi^2_\nu = \chi^2_{min}/\nu$, where $\nu$\ is the number of degrees of 
freedom, gives a quantitative measure of how well the profiles at different 
epochs coincide.  Here $\nu = n - 3$, the number of bins considered, less the 
number of free parameters.  (In addition to the ``interesting'' parameter 
$l$, the sky level and the amplitude scaling factor also represent 
parameters that are adjusted; in practice $n \gg 3$, so this detail is 
insignificant.)  For large $\nu$, one expects $\chi^2_\nu \sim 1$\ 
if the expanding filaments remain essentially constant in shape over the 
span of our observations.   

To calculate the displacements, we chose the 1998 image as the reference, 
and  took 200 bins (a 20\arcsec\ distance), roughly centered on 
the filament peak, from the profile for each sector.  We did the  
comparison of the 1987 and 1991 profiles with the 1998 one for the same 
sector by, in effect, sliding a moving 200-bin window along the earlier 
profile and calculating  $\chi^2(l)$\ by Eq. 2\@.  For each profile pair we  
took the five $\chi^2(l)$\ values closest to the minimum and fitted a 
parabola, in order to obtain a displacement measurement that is not limited 
by the granularity of the 0\farcs 1 bins.  


\begin{table*}

\center{
{\scriptsize 

{\scshape  
\vspace {0.02in}
Table 2 \\

Proper Motion Measurements for the NW Balmer Filaments in SN~1006.} \\

\begin{tabular}{ccccccccccccc}

\\
\tableline \tableline
\\
\vspace {0.04in}

 &  & \multicolumn{4}{c}{1987--1998 Measurement$^{\rm a}$} &  &
\multicolumn{4}{c}{1991--1998 Measurement$^{\rm b}$} &  & 
Mean$^{\rm c}$ \\  

\cline{3-6} \cline{8-11} \cline{13-13}\\

 &   & Correlation & $\chi^2$\ Shift$^{\rm e}$ & 
$(\chi^2_\nu)_{\rm min}$ &$\mu ^{\rm f}$  &  & 
Correlation & $\chi^2$\ Shift$^{\rm e}$  &
$(\chi^2_\nu)_{\rm min}$ &$\mu ^{\rm f}$  & &
$<\mu>$ \\

\vspace {0.04in}
Sector & Azimuth   & Shift$^{\rm d}$\ (\arcsec) &  (\arcsec) & & 
$({\rm mas\; yr}^{-1})$  & &
Shift$^{\rm d}$\ (\arcsec) &  (\arcsec) & & 
$({\rm mas\; yr}^{-1})$  & & 
$({\rm mas\; yr}^{-1})$ \\

\vspace {0.04in}
(1) & (2) & (3) & (4) & (5) & (6)  &  & (7) &  (8) & (9) & (10)  &  & (11)

\\

\tableline \\
A &314\degr--316\degr & &  &  &  &  &
$2.068 $ & $2.070 (15)$ & 1.20 & $288.2\pm 2.1$ & & 
$288.2\pm 2.1$ \\
B & 316\degr--318\degr & 
$3.200$ & $3.209 (19)$ & 1.89 & $287.1\pm 1.7 $& &
$2.036$ & $2.038 (16)$ & 1.04 & $283.7\pm 2.2 $& &
$285.2 \pm 1.4$ \\
C & 318\degr--320\degr &
$3.118$ & $3.180 (32)$ & 1.50 & $282.2\pm 2.9 $& &
$2.044$ & $2.081 (28)$ & 1.89 & $287.3\pm 3.9 $& &
$284.0 \pm 2.3$ \\
D & 320\degr--322\degr &
$3.091$ & $3.087 (21)$ & 0.66 & $276.8\pm 1.9 $& &
$1.935$ & $1.934 (19)$ & 1.29 & $269.4\pm 2.6 $& &
$274.3 \pm 1.5$ \\
E & 322\degr--324\degr &
$3.126$ & $3.128 (18)$ & 1.78 & $280.2\pm 1.6 $& &
$1.991$ & $1.992 (19)$ & 0.72 & $277.4\pm 2.6 $& &
$279.4 \pm 1.4$ \\
F & 324\degr--326\degr &
$3.156$ & $3.156 (13)$ & 0.62 & $282.8\pm 1.1 $& &
$2.040$ & $2.042 (12)$ & 0.53 & $284.2\pm 1.7 $& &
$283.2 \pm 1.0$ \\
G & 326\degr--328\degr &
$3.170$ & $3.172 (16)$ & 1.03 & $284.1\pm 1.4 $& &
$2.029$ & $2.034 (13)$ & 0.89 & $282.9\pm 1.8 $& &
$283.7 \pm 1.1$ \\
H & 328\degr--330\degr &
$3.216$ & $3.219 (18)$ & 0.83 & $288.3\pm 1.6 $& &
$2.038$ & $2.042 (18)$ & 0.82 & $284.1\pm 2.5 $& &
$287.1\pm 1.4$ \\
I & 330\degr--332\degr &
$3.079$ & $3.095 (22)$ & 0.83 & $276.6\pm 2.0 $& &
$1.982$ & $2.001 (21)$ & 0.83 & $277.4\pm 2.9 $& &
$276.8 \pm 1.6$ \\
J & 332\degr--334\degr &
 &  &  & & &
$1.999$ & $1.999 (17)$ & 0.71 & $278.4\pm 2.4$& &
$278.4 \pm 2.4$ \\
\vspace {0.05in}
K & 334\degr--336\degr &
 &  &  & & &
$1.952$ & $1.957 (16)$ & 1.00 & $272.2\pm 2.2 $& &
$272.2 \pm 2.2$ \\

\tableline  \\


\multicolumn{13}{l}{$^{\rm a}$11.16-year baseline.}\\
\multicolumn{13}{l}{$^{\rm b}$7.18-year baseline.}\\
\multicolumn{13}{l}{$^{\rm c}$Weighted mean of the 1987--98 and 1987--98 
measurements.}\\
\multicolumn{13}{l}{$^{\rm d}$Shift in filament position between epochs measured by the 
correlation method.}\\
\multicolumn{13}{l}{$^{\rm e}$Shift in filament position between epochs measured by 
minimizing $\chi^2$; uncertainty in the last two digits shown in 
parentheses.}\\
\multicolumn{13}{l}{$^{\rm f}$Mean of measurements by the correlation and 
minimum-$\chi^2$\ methods, divided by the baseline.}\\

\end{tabular}
}
}

\end{table*}

\subsection{Results}
The results from these measurements are summarized in 
Table 2\@.  Individual values for the shift in filament position, 
as measured both by the correlation method and by minimizing 
$\chi^2$, are given for each sector.   In every case the two methods give 
virtually identical results.   The values for the proper motion, $\mu$\ 
(columns 6 and 10), 
use a simple average of the shifts obtained by the two methods, divided by 
the baseline between epochs.  Measurements are presented for both the 
1987--1998 and 1991--1998 combinations; the third pair, 1987--1991, is not 
independent of the other two and, since it represents the shortest baseline, 
is omitted.  

\begin{figure*}[t]

\plotone{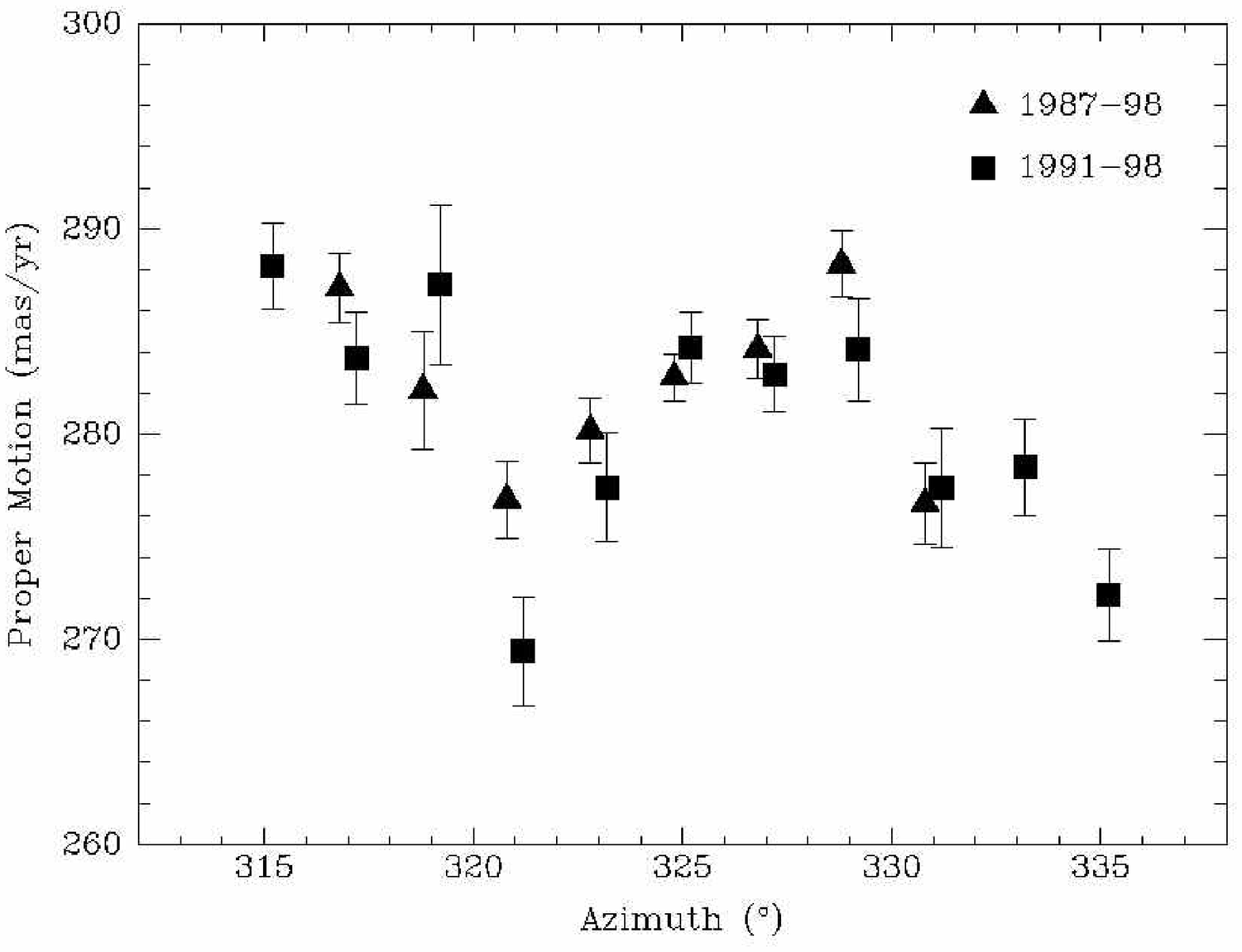}
\caption{Proper motions of the SN~1006 filaments, measured in 
2\degr\ azimuthal sectors, over baselines 1987--98 (triangles) and 
1991--98 (squares).  Points for the two baselines at the same azimuth are 
displaced horizontally for clarity.
}
\end{figure*}

The uncertainties in measured position shifts, columns 4 and 8 of 
Table 2, were estimated by using the
$\chi^2_{min} + 1$\ criterion for a single interesting parameter, as 
discussed in the preceding section.  The same values are carried 
over to give the uncertainties in $\mu$, columns 6 and 10.
The minimum-$\chi^2$\ values per degree of freedom, $\chi^2_\nu$ (columns 5 
and 9), are 
close to 1, as expected provided that the morphology of the filaments has 
not changed appreciably over the period of our observations and that we have properly 
characterized the data.  This, in turn, lends confidence that our 
uncertainty estimates are at least approximately correct.

\begin{table*}
\center{
\scriptsize
{\scshape  
\vspace {0.02in}
Table 3 \\

Wide-Field Imaging Observations of SN~1006.} \\

\begin{tabular}{lcccccc}

\\
\tableline \tableline
\\
\vspace {0.02in}
 &  &  & Scale & \multicolumn{2}{c}{Filter} &   
Exposure   \\
 
\vspace {0.05in}
Date & Telescope   & CCD   &  
($\arcsec\;{\rm pixel}^{-1}$) &
$\lambda_0$\ (\AA) &$\Delta\lambda$\ (\AA)  & 
(s) 

\\


\tableline  \\

1998 Jun 18-20 &CTIO Schmidt &Tek $2048\times 2048$ & 2.32 &6567  & 29 & $16 
\times 600 $ \\
1998 Jun 18-20 &CTIO Schmidt &Tek $2048\times 2048$ & 2.32 &6852  & 95 & $16 
\times 180 $ \\
1998 Jun 24 &CTIO Schmidt &Tek $2048\times 2048$ & 2.32 &6557  & 25 & $13 
\times 600 $ \\
\vspace {0.05in}
1998 Jun 24 &CTIO Schmidt &Tek $2048\times 2048$ & 2.32 &6852  & 95 & $13 
\times 180 $ \\

\tableline


\end{tabular}
}

\end{table*}

A second criterion that indicates that our uncertainty estimates are 
reasonable is comparison between the two measurements of $\mu$\ over 
independent baselines, 1987-98 and 1991-98.  
In almost all of the eight sectors where we have measurements over two 
independent baselines, the two values for $\mu$\ are consistent, within 
the uncertainties.  These results are illustrated graphically in Fig. 4\@.  
The values for the mean proper motion, $<\mu>$\ (column 11 of Table 2), in 
each sector are 
obtained from a weighted (inversely as the variance) mean of the 
measurements over the two independent baselines.

As one moves along the filaments in azimuth, there is some evidence 
for  a variation in the rate of expansion, as shown in Figure 4.  What is most apparent, however, 
is that any such variation is slight, if present at all.  Over the azimuth 
range 314\degr--336\degr, all the measurements lie within the range 
$280\pm 8\ {\rm mas\; yr}^{-1}$, a variation of $< 3\%$.  The new 
measurements all agree with the mean value of 
$300\pm 40\ {\rm mas\; yr}^{-1}$\ reported by  
LBV88 for 10 positions along the filaments---with an
azimuth range very similar to the one where our measurements were made.  
Although LBV88 reported proper-motion values ranging from  
220 to 330 ${\rm mas\; yr}^{-1}$, all but one of their 
ten measurements were consistent (within $\pm 1\sigma$) with no variation 
along the filaments.  
The only other previous measurement is that by \citet{hes81}, who found a 
mean proper motion of $390\pm 60\ {\rm mas\; yr}^{-1}$, but with large 
scatter among their 11 individual values:  140 to 850 ${\rm mas\; 
yr}^{-1}$.  The \citet{hes81} measurement was made by scanning plates 
from two epochs perpendicular to the 
direction of the filaments with a PDS microdensitometer 
and comparing the profiles visually.  This measurement suffered from the 
short (5 yr) baseline between epochs and from poor seeing on the second-epoch 
plate.  The LBV88  measurement was made by digitizing the first-epoch 
image, aligning the images from the two epochs, and then simply measuring 
the position of the filament on the screen using an image cursor.  
We are confident that our measurement, based on CCD images from 
three different epochs, all with excellent seeing, and using two different  but consistent digital  
techniques to measure the shifts, has produced a much more precise 
result.  

\section {Deep Imaging: Faint, Diffuse \ha\ Emission}
In addition to the prominent Balmer-dominated filaments in the NW of 
SN~1006, there is also much fainter,  more diffuse, Balmer emission 
surrounding almost the entire shell and at some places in the interior.  
Discovered in CCD images obtained from 
the CTIO 0.6-m Curtis Schmidt telescope, this emission was reported in WL97.  
In the same 1998 CTIO observing run 
where we obtained the 0.9-m images for the proper-motion study, we also 
obtained a long series of images from the Schmidt, in order to reveal the 
full extent of the faint emission.  Images were obtained through two 
different \ha\ filters (a backup \ha\ filter was used for the 18-20 June 
observations since 
our primary filter was in use at the 0.9-m) and through a 
matched red continuum filter, to facilitate subtraction of the  stars.  
The observational details are given in Table 3.  

Figure 5 shows the \ha\ image after 
continuum subtraction.  Continuum subtraction was 
done separately for the 18-20 June and the 24 June data, and then the 
subtracted images were combined.  The latter image shows, in addition to 
the ``bright'' NW filaments, a more diffuse and much fainter outline of 
\ha\ emission extending around almost the entire shell of SN~1006.  Also 
evident are many features, mostly diffuse emission but with a few 
well-defined filaments, within the shell (as seen in 
projection).  Some of these faint features were evident in Figures 1b and 1c of 
WL97, but the new images reveal far more detail.  Although the 
present images and those from 1993 shown in WL97 were both taken from 
the CTIO Schmidt telescope, the improvement results because the 
present images represent more than 3 times 
longer effective exposure in \ha, and more than 6 times longer in 
the continuum, with a CCD chip that is more sensitive than the one used 
in 1993.  Flux calibration, achieved through observations of several 
spectrophotometric standard stars from the list of \citet{ham92}, yields a 
maximum surface brightness for the NW \ha\ filaments of $2.0 \times 10^{-16}\FLUXARCSEC$, a 
value consistent with the spectrophotometry of \citet{gha02} when their 
spectrum  is folded through our filter.   The more diffuse emission that 
clearly outlines the rim of the shell to the S has surface brightness that 
is lower by almost a factor of 20, 
$1\!-\!1.5 \times 10^{-17}\FLUXARCSEC$.
The large, diffuse oval that is evident within the north central portion of 
the shell in Fig. 5 is even fainter:  $\sim 0.6-1.4 \times 
10^{-17}\FLUXARCSEC \approx 1\!-\!2.5\; {\rm R}$\ 
(1 Rayleigh = $10^6\:{\rm photons}/4\pi \:{\rm sr}$) above the nearby 
background sky.  For comparison, the intensity of the moonless sky within  
our 25 \AA\ bandpass  at the 
time of our observations was $70\!-\!100\; {\rm R}$. 

\begin{figure*}
\epsscale{1.2}
\plotone{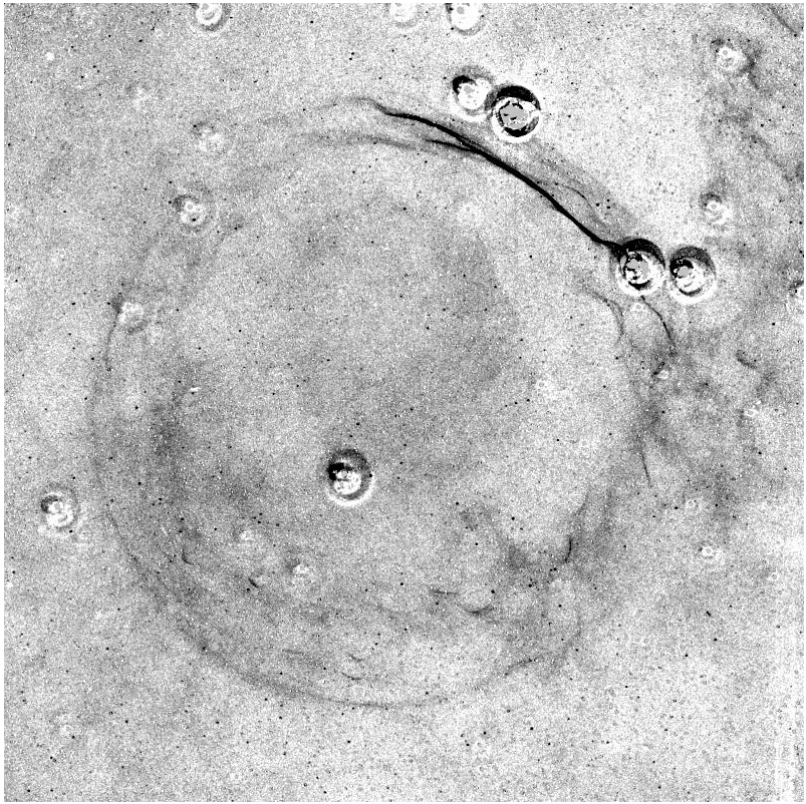}
\caption{Deep \ha\ image of SN~1006 after a matched continuum image has 
been scaled and subtracted.  In this display, the NW filaments are badly
saturated in order to reveal the faint emission that outlines virtually the entire 
shell of the remnant.  The characteristic pattern around all the bright 
stars is an artifact, probably resulting from microscopic imperfections in 
the filters.  The field is exactly 40\arcmin\ square; N is up and E to the 
left. 
}
\end{figure*}
 
There is no doubt that the features shown in Figure 5
are real and not artifacts; all are evident in continuum-subtracted images from each of
the two individual \ha\ filters, while deep images of other fields show no such
structures.  In particular, the images show that although the remnant shell is well
defined at most azimuthal angles, there is diffuse emission extending 
inward $\gtrsim 1/4$\ of the radius.   Since the lifetime of the 
neutral H behind the shock is quite short, it is extremely unlikely that 
the interior emission stems from H atoms that have penetrated that far 
behind the shock front; instead we must be 
observing material projected along the line of sight. The diffuse shell is
particularly evident in the SW, where radio maps also reveal several ribbons of emission.
Furthermore, the outer edge of the SNR in the NW extends well beyond the bright optical
filaments, making the entire SNR more spherical in appearance and supporting arguments
that the bright filaments have arisen where the SNR shock has 
impacted a localized region of greater than average density.  

Fig. 5 also shows diffuse emission that extends far beyond the radio 
or X-ray boundary of the SNR to the W and NW.  
Emission extends in these directions well beyond the field shown in Fig. 5 
and is almost certainly diffuse foreground or background Galactic emission 
unrelated to SN~1006.  
The intensity of the diffuse emission is 
0.6--2 $\times 10^{-17}\FLUXARCSEC$, or 1--3 R, which corresponds 
to emission measure $\approx 2-6\:{\rm cm}^{-6}\:{\rm pc}$.  Some of the emission 
within the (projected) SN~1006 shell, especially in the SW, may be an extension of this 
Galactic emission.  
\citet{dub02} have conducted a search for \ion{H}{1}  21-cm emission in the 
direction of SN~1006 using the Australia
Telescope Compact Array.  Their images, which have a resolution of about 4 arcmin, show
a number of concentrations around the SNR, and it is possible that portions of these
features also appear in our \ha\ images. Their total \ion{H}{1} column map
also shows an increase near the center of the SNR, which could possibly be associated with
the diffuse \ha\ emission covering the central region of the shell. 
However, it is entirely reasonable that the central \ha\ emission 
is associated with SN~1006 itself, resulting from a face-on shock 
at the front or rear of the shell.  
An average of $\sim 0.05$\ \ha\ photons are produced for every neutral H 
atom entering a shock \citep{che78}, so a $3000 \kms$\ shock propagating 
along the line of sight
into a region with $n_{\rm H\:I} = 0.1\: {\rm cm^{-3}}$\ would produce  
a diffuse intensity of $\sim 1.5 \: {\rm R}$\@.
In the absence of spectroscopy, which we would expect 
to show Balmer lines with both narrow and very broad components for 
emission from the nonradiative shocks that characterize SN~1006 itself
(versus only narrow lines from diffuse Galactic emission), it is impossible 
to resolve this issue.  Spectroscopy of material within the (projected) 
interior that is truly associated with SN~1006 shocks 
would be interesting and would help to elucidate the three-dimensional 
geometry of the remnant, but measuring the width and center of broad Balmer 
lines from such faint, diffuse material will present a serious 
observational challenge.

We note in passing that very little if any of the emission in the image is of synchrotron
origin, contrary to the suggestion by \citet{all01} about our earlier 
(WL97) H$\alpha$ image.  Synchrotron emission should appear stronger in the 
matched continuum images, which have a broader bandwidth, than it does in 
the \ha\ ones.  Instead, no diffuse emission whatsoever is evident in our 
continuum images, and the diffuse emission---barely visible in the 
\ha\ image before subtraction---becomes evident in the difference image between \ha\ and the 
matched continuum (Fig. 5).  
Furthermore, the diffuse \ha\ 
emission does not follow the radio or X-ray images well, contrary to what 
one would expect if all are of synchrotron origin.

\section{Expansion rates and Azimuthal Variation}

The expansion of  SNRs at various stages of their 
development can be expressed in idealized, spherically symmetric 
approximations as power laws for the outer shock radius  
as a function of time, $R \propto t^m$, where $m$\ is an expansion index.   
\citep[\eg][]{wol72,ost88,mof93,jon98}.  
Very early, when the ejecta dominate over 
swept-up material, the ejecta expand freely and  $R \propto 
t$\@.  As the ejecta begin to encounter interstellar and circumstellar 
material, a reverse shock develops---stages for which \citet{che82}\ has 
obtained similarity solutions that give $0.8 \gtrsim m \gtrsim 0.5$, 
depending on the density dependence of the ejecta and the 
circumstellar material.  As the reverse shock approaches the  
center, it disappears; the remnant becomes dominated by swept-up matter; 
and the expansion is described by the familiar 
Sedov (adiabatic) phase,  $R \propto t^{2/5}$,  $m = 0.4$.  
Once the shock slows to a velocity $\sim 200 \kms$, post-shock material 
can cool efficiently and a ``pressure-driven snowplow'' develops, for 
which $m = 0.29$.  Eventually, cooling reaches the 
interior,  the pressure drops, and the expansion approaches the simple 
snowplow, with $m = 0.25$ as dictated by conservation of momentum.  

For SNRs of known age $t$, the mean expansion index, 
$m={{{{dR} \over {dt}}} \mathord{\left/ {\vphantom {{{{dR} \over {dt}}} 
{{R \over t}}}} \right. \kern-\nulldelimiterspace} {{R \over t}}}=
{{\mu t} \mathord{\left/ {\vphantom {{\mu t} \theta }} \right. 
\kern-\nulldelimiterspace} \theta }$, where $\theta$\ is the angular radius 
to the filament whose proper motion, $\mu$, has been measured,  
gives an indication of the evolutionary state of the remnant. 
\citet{mof93} measured the overall expansion of the radio shell at 
frequencies of 1370 and 1665 GHz over a baseline of 11 years 
and found $m = 0.48 \pm 0.13$.  
This suggests a remnant late in its reverse-shock phase and making the 
transition to an ISM-dominated Sedov blast wave.  Since there are no sharp 
radio features along the NW rim of the shell where the Balmer filaments are 
located, Moffett et al. were not able 
to measure the proper motion there and make a direct comparison with the 
optical value---at the time $m = 0.33 \pm 0.04$\ measured by LBV88.  

Our more precise measurement of the optical proper motion leads to a 
refinement of the LBV88 value.  
For the position along the NW filaments, sectors F \& G where our spectra 
were obtained, we measured $\mu = 281 \pm 5 \masy$.  To calculate the 
corresponding radius, we use the apparent center of the faint 
optical emission that extends almost 180\degr\ around the rim from NE 
counterclockwise to SW: 
$\alpha(2000) = 15^{\rm h}02^{\rm m}55\farcs4,\ \delta (2000) = 
-41\degr 56\arcmin 33\arcsec$, a point $\sim 40\arcsec$\ east of the radio 
center used by \citet{mof93}.  From our optical center, the angular radius 
$\theta = 13\farcm 6$, and we find $m=0.34\pm 0.01$.   This value suggests that 
in the NW portion of the shell the blast wave has encountered enough material 
that it is already moving beyond the Sedov phase.  
If the blast wave has encountered a localized,  relatively dense region 
here, it would have slowed significantly compared with its velocity 
elsewhere around the shell.  This in turn would produce both a lower value for 
$m$\ and a smaller radius to the NW filaments than the $\theta \approx 
15\arcmin$\ that characterizes most of the shell---both consistent with 
the observations.  
In their \ion{H}{1} maps, \citet{dub02} have identified 
enhancements, just outside the SN~1006 shell to the NW, that may be interacting 
with the expanding shock to produce the bright filaments.  The velocities 
of these enhancements are $\sim -6 \kms$\ and $\sim + 10 \kms$, rather 
different from the $\sim -25 \kms$\ predicted for material  at the 
distance to SN~1006 based on Galactic rotation models.  As Dubner et 
al. point out, however, one must use caution in applying circular rotation models 
for sources far from the Galactic plane.

It would be interesting to make a direct comparison of optical and 
radio proper motions around the S and E portions of the shell.  Eventually 
it should be possible to measure the expansion of the faint shell of 
Balmer emission (Fig. 5), but we will probably have to wait another decade to 
obtain a baseline long enough to measure the proper motion of this faint 
optical emission, which is far less crisp  than the NW filaments that are 
the subject of the measurements reported in Section 3.

\section {Distance to SN~1006}
The prominent NW filaments in SN~1006 almost certainly represent 
sheets of material seen nearly edge-on, excited by a nonradiative shock 
that is moving outward almost transverse to our line of sight.  
From the shock velocity, $v_s$, and the proper motion, $\mu$, of the filament, the 
distance to SN~1006 is geometrically determined, 
$$d=2.1095\;{\rm kpc}\left( {{{v_s} \over {3000\kms}}} \right)\left( 
{{{300\masy} \over \mu }} \right).$$
The width of the broad component of the Balmer lines, which results from neutral H atoms 
that have penetrated the shock and undergone charge exchange with hot 
protons, is directly  proportional to the postshock proton temperature.  However, 
the rate of electron-proton 
equilibration behind the shock and the pre-shock ionization structure 
introduce dependencies in obtaining the shock velocity.  \citet{gha02} 
show that the ratio of broad to narrow \ha\ flux, together with the strengths of 
(very faint) \ion{He}{1} and \ion{He}{2} lines, indicates minimal 
equilibration at the shock front, and they obtain  an unambiguous 
determination of the shock velocity: $v_s=2890 \pm 100 \kms$.  Furthermore, the near 
coincidence in central wavelengths for the broad and narrow \ha\ 
components in the \citet{gha02} spectrum demonstrates explicitly 
that the shock motion is perpendicular  to the line of sight, 
within $\leq 2\degr$.

The \citet{gha02} spectrum was taken over a 
portion of the filament $ 51\arcsec$\ long, located in sectors E and 
F of Fig. 2 (compare Fig. 1 of \citet{gha02}).  
From Table 2, the measured proper motion of the 
filament in these two sectors is $279.4\pm 1.4$\ and $283.2 \pm 1.0\;\masy$, respectively.  For 
the proper motion of the region corresponding to the $v_s$\ measurement, 
we have used a simple mean, $\mu = 281\pm 5\masy$, where the uncertainty 
is chosen conservatively to embrace some two/thirds of  our 
measurements for {\it all} sectors and to allow for systematic effects.  Combining the 
measurements leads to a geometric distance to SN~1006 of $2.17\pm 0.08\; 
{\rm kpc}$.  The overall precision of $4\%$\ is limited primarily by the 
uncertainty in deriving the shock velocity from the measured widths of the 
broad Balmer lines.  

Our distance estimate is consistent with, but an improvement upon, 
previous measurements 
by the same method:  1.7--3.1 kpc (LBV88); 1.4--2.8 kpc \citep{smi91}; 
$1.8\pm 0.3$\ kpc \citep{lam96}.  
Ours is an improvement both because it is based on the more sensitive digital 
image data presented here and on the superior spectra of \citet{gha02}, and 
because it reflects our improved understanding of the physics of 
nonradiative shocks in SN~1006.  In particular, LBV88 and \citet{smi91} 
allowed the possibility of full equilibration of electrons and ions 
behind the shock, which led to 
a significantly higher $v_s$\ and hence a greater distance.  However, 
\citet{lam96} and \citet{gha02} have showed that equilibration at the shock 
front is minimal and hence the higher shock velocities are excluded.

Barring some fundamental misunderstanding of the physics of  
nonradiative shocks, our geometry-based measurement is 
the most reliable and precise determination of the distance to SN~1006. 
There are, however, other relevant data for constraining this distance. 
UV spectra of the 
\citet[][S-M]{sch80} star, a very blue subdwarf star located near the 
projected center of SN~1006, show strong, very broad absorption lines of \ion{Fe}{2}, 
\ion{Si}{2}, \ion{Si}{4} and other ions, the result of absorption by cold, rapidly 
expanding SN ejecta  within the SN~1006 shell \citep{wu83,wu97}.  Obviously the S-M star must 
be located behind SN~1006, which sets an upper limit on the distance to the 
remnant.   \citet{sch80} classified the star as an sdOB, with color 
excess $E(B-V)=0.11$, to obtain a distance of 0.5--2.5 kpc to the S-M 
star.  \citet{fes88} argued that the S-M star (and sdOB stars in general)  
is probably more luminous than \citet{sch80} had estimated and obtained a 
distance of 1.5--3.3 kpc.  
More recently, \citet{bur00} classified the star as sdB, somewhat cooler but 
more highly absorbed with $E(B-V)=0.16$, to obtain a distance estimate of 
1.05--2.1 kpc.  All of these estimates for the distance to the S-M star are 
consistent, just, with its being behind the SNR.  But it is clear that a 
better understanding of the nature of the S-M star itself is required before 
estimates of its distance can be relied upon.  It may be that  
SN~1006 and the S-M star are physically quite close together, 
fueling speculation that the association is more than coincidental.  
\citet{sav82} have argued convincingly that the S-M star could 
not have cooled fast enough to be the stellar remnant of a SN that occurred only 
1000 years ago, but \citet{bur00} have speculated that it could be the 
stellar remnant of the donor star in an interacting binary system that 
produced an SN~Ia event in 1006 C.E.   

A second line of argument for the distance to SN~1006 also involves the S-M 
star, this time as the light bulb against which to observe broad UV 
absorption lines---and thus to obtain a ``core sample'' of the expanding cold 
material within the SN~1006 shell.  \citet{wu93} reported \ion{Fe}{2} 
velocity profiles extending to $\pm8300\kms$\ at the continuum, indicating 
that material expanding this rapidly must still be contained within the remnant 
shell.  Subsequently, \citet{ham97} reanalyzed the same spectra, along 
with new data \citep{wu97}, and concluded that the \ion{Fe}{2} velocity is lower, and 
that the fastest material is \ion{Si}{2}, for which they observed 
velocities up to $7070\kms$.  Material expanding at the 
latter rate for 987 years (the age of the remnant at the time of the 
observations) will reach a radius of 7.1 pc (or 8.4 pc for the $8300 
\kms$\ velocity).  If we assume spherical symmetry, SN~1006 must be at a minimum 
distance of 1.6 kpc (1.9 kpc)   to contain this material with the 
15\arcmin\  radius of the remnant shell.  

By a similar argument, the {\it maximum} velocity of supernova ejecta might 
provide an upper limit on the distance.   Certainly the present radius of the SNR shell 
cannot exceed the distance the fastest SN ejecta would have covered had there 
been no deceleration.  Several authors \citep[\eg][]{gre84, fes88} have used a maximum velocity of 
$\sim 10,000\kms$\ to obtain an upper limit on the distance, 
$d \lesssim 2.3$\ kpc.  Recent observations \citep[\eg][]{hat99} 
and theory \citep[\eg][]{hof98} show that velocities of 
$16,000-20,000\kms$, and sometimes higher, are found in SN~Ia ejecta.  
The requirement that the average expansion velocity of the shell be 
$\lesssim 20,000\kms$\ requires only that $d \lesssim 4.6$\ kpc---no longer 
a meaningful constraint.  

The distance to SN~1006 has been recently discussed by several authors  
\citep[][WL97]{str88, wil96, sch96, lam96}.  In addition to the 
kinematic arguments discussed above, other distance estimates to SN~1006 
have relied heavily on models for the 
X-ray emission and parameters such as the post-shock temperature and interstellar density that are 
highly uncertain and variable around the remnant shell.  The rich 
X-ray structure seen in recent {\it Chandra} images \citep{lon02} make it 
obvious that earlier models, most of which rely on simple spectra and spherical 
symmetry, are vastly oversimplified.  
We conclude that the measurement of $2.17\pm 0.08\; {\rm kpc}$\ presented 
here is not only consistent with all present data, but is also both more 
precise and more accurate than previous distance estimates.

\section{How bright was SN~1006?}
The supernova that was first witnessed on or about 1006 May 1 is generally 
acknowledged to have become (within a few days) the brightest supernova 
recorded in human 
history.  Despite its southerly declination, 
$\delta(1006) \approx -38\fdg 5$, clear records of its sudden appearance 
are found in 
contemporary chronicles from  Egypt, Iraq, Italy, Switzerland, 
China, and Japan, with additional references that may refer to sightings 
of the star from France, Syria, and elsewhere.  
Interesting excursions through the historical records and 
their interpretation may be 
found in \citet{shk68}, \citet{ste77}, \citet{cla77}, \citet{dev85}, and 
references in these sources.
Recorded observations are much more numerous and 
widespread than for the far 
more favorably positioned (for northern observers) SN~1054 that 
occurred only 48 years later and 
that produced the Crab Nebula.   The peak magnitude for SN~1006 
has been variously estimated at values ranging from 
$\sim -5\ {\rm to}\ -10$, based on different interpretations of the 
historical accounts (see below).




Once upon a time, astronomers sought to shed light on the Hubble constant 
by combining the brightness at maximum of SN~1006 and of other historical 
supernovae with the modern distances to their remnants to estimate peak 
absolute magnitudes.  These estimates then served as a method for obtaining the 
distances to extragalactic SNe.  Although recorded sightings of 
SN~1006 by eleventh-century observers are numerous, photometric 
``calibration'' of their observations 
has proved fraught with ambiguity for modern readers, 
as the wide variation in interpretation of the same records attests.    
Fortunately, the Cepheid distance scale has now been extended 
to include the host galaxies of 
several well-observed supernovae with good photometry.   We can thus run the 
old argument in reverse, using current values for SN peak absolute 
magnitudes and the distance to SN~1006 to estimate just how bright it was 
at maximum.  \citet{lam96} have followed an argument similar to the one 
below, but with  different, more uncertain, values both for the distance 
to SN~1006 and for its absolute magnitude at maximum.   

While we cannot be certain of the type  for SN~1006, it is generally 
assumed to have been Type Ia.  Arguments include its location, at Galactic 
latitude $14\fdg 6$, $\sim 550$\ pc above the plane, the presence of at 
least a few tenths $M_{\sun}$\ of Fe within its shell \citep{ham97}, the 
absence of any evidence for a compact stellar remnant, the lack of any 
massive star association anywhere in the vicinity, and the implication from 
at least one Chinese historical record, the {\it Sung Shih}, official 
history of the Sung dynasty (960--1279), that it remained visible for several years 
\citep{go_ho65}\@.  \citet{sah01} give 
peak absolute magnitudes for 9 SNe~Ia in galaxies with well-determined 
Cepheid distances, and find the mean V-band absolute magnitude at maximum 
light (uncorrected for decline rate) to be $M_V^{max} = -19.55$, with an RMS 
dispersion of 0.09\@.  Both \citet{bra98} and \citet{gib00} have used almost 
the same data set with somewhat different calibration procedures to 
arrive at mean absolute magnitudes of -19.44 and -19.47, respectively, and 
a dispersion of 0.12 in the latter case.
A fairer representation of the actual 
dispersion in peak luminosities of SNe~Ia is probably that based on 
a sample of 29 uniformly observed SNe~Ia in the the 
Calan-Tololo survey \citep{ham96}, for which $\Delta M_V^{max} = 0.26$.  
Correction based on the rate of decline \citep{phi93} would reduce the 
dispersion, but the 
decline rate is of course not known for SN~1006.  We will adopt here a 
peak luminosity for SN~Ia events corresponding to $M_V^{max} = -19.50 \pm 0.40$, 
where the uncertainty should be generous enough to include both uncertainties 
in the calibration and the intrinsic 
dispersion among the events themselves.

The reddening to the S-M star was estimated by \citet{sch80} as 
$E(B-V)=0.112 \pm 0.024$, and the visual extinction as $A_V = 0.32\pm 0.10$. 
\citet{wu93} used the UV spectrum of the S-M star to obtain essentially the 
same value:  $E(B-V)=0.10 \pm 0.02$, corresponding to $A_V = 0.31\pm 0.06$.
Since the star must be beyond, but probably not far beyond, SN~1006, we  
adopt the same extinction value for the remnant with somewhat higher 
uncertainty: $A_V = 0.31\pm 0.10$.  

Assuming that SN~1006 was, in fact, a Type Ia event and using the above 
values for peak SN~Ia luminosity, and the distance and extinction to SN~1006, we 
conclude, {\it a posteriori}, that at its brightest, SN~1006  
reached peak visual magnitude $-7.5 \pm 0.4$\@.  The uncertainty is a purely 
formal one, and is dominated by that in the peak SN~Ia luminosity.
  
Our value falls squarely in the middle of the widely varying estimates based on historical 
records.   Most of these have relied heavily on the 
same  text by the Egyptian astrologer Ali bin Ridwan (d. 1061), 
in a commentary on Ptolemy's {\it 
Tetrabiblos} rediscovered  by \citet{gol65}, of the remarkable 
stellar  spectacle that Ali bin Ridwan recalled from his youth.   
As translated from the Arabic by Goldstein, this reads in part, 

\begin{quote}
I will now describe for you a spectacle that I saw at the beginning 
of my education. \ldots ({\it he describes the location, in opposition to the 
Sun}) \ldots\ It was a large spectacle, round in shape, and its size 
2\onehalf\ or 3 
times the magnitude of Venus.  Its light illuminated the horizon and it 
twinkled very much.  The magnitude of its brightness was a little more than 
a quarter of the brightness of the Moon.  \ldots
\end{quote}

This passage has been variously interpreted to arrive at quantitative 
estimates for the magnitude of SN~1006 at maximum.  \citet{gol65} 
focused on comparison with the Moon and reasoned that the peak brightness 
was comparable to that of the Moon quarter-illuminated (--8) or 
half-illuminated (quarter phase, --10).  
\citet{ste77} argue that the comparison should be made with the full Moon.   
Changing light levels have a subjective effect approximating the square 
root of the intensity,  so the maximum brightness of the star would have 
been about 1/15 that of the full Moon, or magnitude --9.5\@.  (Stephenson 
et al.  also invoke other arguments based on this and other texts to 
support the same estimate.) 
\citet{psk78} argued that the star must have been 
2\onehalf\ or 3 magnitudes brighter than Venus and used the modern 
magnitude scale to arrive at $-6 \pm 0.5$\@.   Independent of the 
Ali bin Ridwan text, \citet{sch96} 
has estimated the peak magnitude based on historical records of the 
heliacal setting and rising of SN~1006 in August and November of 1006, 
respectively.  By comparison with a variety of template light curves for 
SNe Ia , he estimates a peak magnitude between --4.1 and --6.7\@. 

There is another, very straightforward interpretation of 
the Ali bin Ridwan  
text that gives a result remarkably similar to our current estimate.  
Suppose that we assume that ``its size 2\onehalf\ or 3 
times the magnitude of Venus,'' and ``the magnitude of its brightness 
was a little more than a quarter of the brightness of the Moon,'' both 
refer to the {\it same} sort of relative measurement, on a logarithmic 
scale analogous to the modern (and also the ancient, approximately) 
magnitude scale.  That is, Venus and the Moon 
represent points of comparison, and the spectacle in question is  
2\onehalf\ or 3 steps brighter than Venus, and 4 steps fainter than the Moon.
On the evening of 1006 May 1, the new star appeared  in opposition to the 
Sun, while the Moon was only one day past new and thus was not visible.  
At the full Moon two weeks later, however, the supernova was probably near 
its peak brightness.  Toward the end of twilight, the full Moon, SN~1006, 
and Venus would all have been visible $\sim 15\degr$\ above the horizon in 
different directions.  It seems entirely plausible that this memorable  
array would have prompted Ali bin Ridwan to make the comparison quoted 
above---using Venus and the Moon as benchmarks to gauge the 
brightness of the star.  At the time, Venus would have had $V \approx -4.0$, and 
the full Moon $V \approx -12.5$\ \citep{ste77}.  
If we divide the 
magnitude range $V = -4.0\ {\rm to}\ -12.5$\ into 6.5 (7) equal steps, and 
take the SN~1006 magnitude as 2.5 (3) steps brighter than Venus, we find a 
result $V_{SN} \approx -7.3\ (-7.6)$.   This interpretation seems at least 
as plausible as any other, and agrees remarkably well with our estimate 
based on modern data.  

While a peak brightness of $-7.5$\ is significantly fainter than the more  
extravagant of past estimates based on historical data, 
certainly an event of magnitude 
$-7.5$\  would have been bright enough to attract wide attention, as 
indeed SN~1006 did throughout the world in the eleventh century.


\acknowledgments

P.F.W. and K.S.L. gratefully acknowledge the outstanding support,  
typical of the mountain staff at CTIO, during the observations that yielded the 
new data reported here.  We thank F. D'Arcangelo for her assistance in 
reduction of the Schmidt data, J. D. Emerson for sharing his insight 
into some  statistical questions, and M. Kamal for translation and 
elucidation of the 
Arabic texts relating to SN~1006.  This work has been made 
possible through the financial support from the NSF, through grant 
AST-961845 to P.F.W., and from NASA, through grants NAG 5-8020 to P.F.W. and 
{\it Chandra} grants GO0-1120X and GO1-2058A to K.S.L.  Additional support for 
astrophysics research at Middlebury College has been provided by the W.M. 
Keck Foundation through the Keck Northeast Astronomy Consortium.




\bibliographystyle{apj}

\bibliography{bibmaster}

\end{document}